\documentclass[reqno]{amsart}
\usepackage{a4}

\usepackage{dsfont}
\usepackage{amscd,color}

\usepackage{graphicx}
\usepackage{srcltx}
\usepackage[english]{babel}
\usepackage{amsmath}
\usepackage{amsthm}
\usepackage{amsmath}
\usepackage{amsfonts}
\usepackage{amssymb}
\usepackage{amscd}
\usepackage{enumerate}
\usepackage{dsfont}
\usepackage{cite}

\usepackage{appendix}

\usepackage{rotating}

\usepackage[automark]{scrpage2}

\def\R{\mathbb{R}}
\def\C{\mathbb{C}}
\def\H{\mathcal{H}}

\def\A{\mathcal{A}}
\def\G{\mathcal{G}}

\def\s{\sigma}
\def\p{\psi}

\def\2{\frac{1}{2}}
\def\ii{\frac{1}{2i}}
\def\Tr{\mathrm{Tr}}

\def\be{\begin{equation}}
\def\ee{\end{equation}}
\def\bp{\begin{proof}}
\def\ep{\end{proof}}
\def\bc{\begin{cases}}
\def\ec{\end{cases}}

\newcommand{\bra}[1]{\ensuremath{\left\langle #1\right|}}
\newcommand{\ket}[1]{\ensuremath{\left|#1\right\rangle}}
\newcommand{\braket}[2]{\ensuremath{\left\langle #1\vphantom{#2}\right.\left|\vphantom{#1}#2\right\rangle}}

\numberwithin{equation}{section}

\newtheorem{Def}{Definition}[section] 
\newtheorem{Pro}[Def]{Proposition}
\newtheorem{The}[Def]{Theorem}
\newtheorem{Cor}[Def]{Corollary}
\newtheorem{Rem}[Def]{Remark}

\begin{document}

\markboth{P. Aniello, J. Clemente-Gallardo, G. Marmo and G. F. Volkert}{Classical Tensors and Quantum Entanglement II: Mixed States}

%
%

\title{CLASSICAL TENSORS AND QUANTUM ENTANGLEMENT II: MIXED STATES}

\author{P. ANIELLO}

\address{Dip. Sc. Fisiche, Fac. Sc. Biotecnologiche and INFN-Napoli, Universit\`a Federico II\\
Via Cintia, Napoli, 80126, Italy
}

\author{J. CLEMENTE-GALLARDO}

\address{BIFI and Departamento de F\'isica Te\'orica\\
Edificio I+D-Campus R\'io Ebro, 50018 Zaragoza, Spain 
}

\author{G. MARMO}

\address{Dip. Sc. Fisiche and INFN-Napoli, Universit\`a Federico II\\
Via Cintia, Napoli, 80126, Italy
}

\author{G. F. VOLKERT}
\address{Dip. Sc. Fisiche and INFN-Napoli, Universit\`a Federico II\\
Via Cintia, Napoli, 80126, Italy\\
and\\
Mathematisches Institut, Ludwigs-Maximilians-Universit\"at\\ 
Theresienstr. 39, 80333 M\"unchen, Germany\\
}


\maketitle

\begin{abstract}
Invariant operator-valued tensor fields on Lie groups are considered. These define classical tensor fields on Lie groups
by evaluating them on a quantum state. This particular construction,
applied on the local unitary group $U(n)\times U(n)$, may establish a
method for the identification of entanglement monotone candidates by
deriving invariant functions from tensors being by construction
invariant under local unitary transformations. In particular, for $n=2$,
we recover the purity and a concurrence related function (Wootters
1998) as a sum of inner products of symmetric and anti-symmetric parts
of the considered tensor fields. Moreover, we identify a distinguished
entanglement monotone candidate by using a non-linear realization of
the Lie algebra of $SU(2)\times SU(2)$. The functional dependence
between the latter quantity and the concurrence is illustrated for a
subclass of mixed states parametrized by two variables.   
\end{abstract}



\section{Introduction}

In  previous papers \cite{Aniello:08,Aniello:09}, we have shown how to construct
classical tensor fields on a Lie group $\G$ out of quantum states and unitary
representations of the Lie group $\G$. This procedure resembles what has
been done by Peremolov in defining generalized coherent states
\cite{Perelomov:1971}. Our approach, however, is an aspect of a general
program for a geometrical formulation of quantum mechanics
\cite{Ashtekar:1997ud,Clemente:2008,Brody:2001,Carinena:2007ws} 
which aims at a description of quantum mechanics in classical differential
geometric terms. In this picture, it is possible to consider as a 
`classical' (finite-dimensional)
manifold $M$ any (finite-dimensional) Hilbert space $\H$, or its manifold of
rays, the complex projective space associated with $\H$, say $R(\H)$
or $\mathbb{CP}(\H)$. Somewhere else \cite{Chruscinski:2008}, it has also been
shown how to deal 
with the GNS construction in this geometrical formulation so that a
highly unified picture  emerges showing that most concepts and
mathematical construction of symplectic and Riemannian geometry are
unified in this 
K\"ahlerian description of quantum mechanics. States and observables are
connected with the Hilbert space description by means of the momentum map
associated with the symplectic action of the unitary group
$U(\H)$. Our construction of tensor fields on the Lie group $\G$  is
to be thought within the general picture of a
quantization-dequantization scheme that has been considered
recently in several papers (see \cite{Ibort:2009} for a review on these works). The latter approach involves the construction of a star product of complex-valued functions on the group
$\G$ induced by the product of operators in the Hilbert space $\H$
carrying a unitary representation $U$ of $\G$ (representation that gives rise to an action of $\G$ on the physical states\footnote{By `physical states' we mean the so-called \emph{normal states} on the algebra of observables, namely, those states that can be identified with density operators. As is well known, in the case of a finite-dimensional quantum system \emph{all} states (i.e.\,all normalized positive functionals on the algebra of observables) are normal states.}). Now, in addition to the `symbol' $f_A\colon \G \rightarrow \C$ associated with the operator $A$, i.e. 
$$
f_{A}(g)=\mathrm{Tr}(AU(g))
$$
--- for sake of simplicity we are assuming that $\H$ is finite-dimensional, or, more in general, $A$ is a trace class operator, and $\G$ is unimodular (see \cite{Aniello2009A}) --- we also consider
$$
df_{A}(g)=\mathrm{Tr}(AdU(g))
$$
and tensor fields which
may be constructed out of them on the group.\\  
Besides paving the way towards an alternative, more general approach to
dequantization, the more concrete justification for these
constructions may turn out to be provided from current questions of
quantum information, being not necessarily connected to any classical
limit. In particular, it has been shown in \cite{Aniello:09} that a
characterization of quantum entanglement for pure states is closely
related to the pull-back tensor field  
\be T_A(g) :=\mathrm{Tr}(AU^{-1}dU(g)\otimes U^{-1}dU(g))\label{ProjectivePBT Intro}\ee 
on the Lie group $\G=U(n)\times U(n)$ arising from a covariant Hermitian tensor field
\be d\bar{z}^{j}\otimes dz^{j}\label{ProjectiveHT Intro}\ee
defined on a finite-dimensional Hilbert space $\H$ whenever the
operator $A$ is considered as a  normalized rank-1 projector, say
$\ket{0}\bra{0}$, defining a pure quantum state.  The pull-back
(\ref{ProjectivePBT Intro}) is induced by a unitary action of the
underlying Lie group $\G$ on a normalized fiducial vector $\ket{0}\in
\H$ \cite{Aniello:08, Aniello:09,Volkert:2010}. Such a pull-back
construction should in principle exist also in the algebraic approach
to quantum mechanics, by starting with a Hermitian tensor field on a
$C^*$-algebra of operators, instead of a Hilbert space. To capture
also an entanglement characterization for \emph{mixed}, rather than only
for pure states, via a corresponding generalized tensor field
construction, we shall take into account a different, probably more
direct route. \\ 
This is done here in the following section \ref{IOVT-section}, where we present a
general framework for a construction of \emph{invariant operator-valued tensor fields} (IOVTs)
\be U^{-1}dU(g)\otimes U^{-1}dU(g)\ee
directly defined on Lie groups. The
corresponding operators may either be
represented here on vector spaces, or more generally, realized on a
manifold. The IOVT-construction will be applied in Section \ref{Evaluating IOVTs on composite systems} on the unitary group $U(n)$ in arbitrary finite dimensions
with an evaluation of these tensors on states associated to $n$-level quantum systems. Thereafter, we focus particular attention on
an entanglement characterization of bi-partite mixed quantum states by
applying our framework of IOVTs on the local unitary group $U(n)\times U(n)$. In section \ref{Applications on subclasses of states in n=2}, we will illustrate our results on particular
families of entangled states for $n=2$.  We end up with conclusions
and an outlook to the Peres-Horodecki criterion in section \ref{outlook}.

\section{Invariant Operator-Valued Tensor fields on Lie groups}\label{IOVT-section}

\subsection{Invariant tensor fields on $\G$}

Given a matrix Lie group $\mathcal{G}$, we may construct a left invariant Lie algebra-valued 1-form 
by setting 
\be g^{-1}dg\equiv \tau_j\otimes\theta^j,\label{1,1}\ee
where $\{\theta_j\}_{j\in J}$ denotes a basis of left-invariant 1-forms on $\mathcal{G}$, and $\tau_j$ is a basis for the Lie algebra of $\G$. By using a dual basis of left-invariant vector fields $\{X_j\}_{j\in J}$, defined by  
\be i_{X_j}(g^{-1}dg) = \tau_j,\ee
we find that the left invariant Lie algebra-valued tensor field (\ref{1,1}) is related to the identity (1,1)-tensor field field $ X_j\otimes\theta^j$.
From the manifold view point, the composition law
\be \varphi: \mathcal{G}\times\mathcal{G}\rightarrow \mathcal{G} \ee
may be considered as an action of $\mathcal{G}$ on itself, and $dg$ denotes the `differentiation' with respect to the action, and not with respect to the `point' which is acted upon.\\ 
The Lie algebra valued 1-form (\ref{1,1}) is known to be a standard construction on Lie groups (see e.g. \cite{Helgason:1962} for further details). The generalization of this construction to higher order tensor fields is usually done in this regard in terms of the wedge, rather then the ordinary tensor product. In contrast to this considerations, here we would like to take into account higher order tensor fields taking values \emph{also in the tensor algebra rather then only in the Lie algebra}. Let us now illustrate this idea. \\
By considering the matrix-valued differential 1-form $dg$ we may construct the tensor field 
\be dg^{-1}\otimes dg = -g^{-1}dg \otimes g^{-1}dg,\ee  
where we have used
\be dg^{-1}= -g^{-1}dg g^{-1}.\ee
The tensor field becomes then
\be -(X_j\otimes\theta^j)\otimes (X_k\otimes\theta^k)\ee 
which may also be written as 
\be -(X_j\otimes X_k)\otimes(\theta^j\otimes \theta^j),\ee
that is a \emph{tensor algebra valued left-invariant tensor field} on $\G$.   

\begin{Rem}
Had we chosen $dg\otimes dg^{-1}$, we would have a similar expression in terms of right-invariant 1-forms and right-invariant vector fields.  
\end{Rem}
 
\begin{Rem}
Higher order tensors may be constructed in similar manner.  
\end{Rem}

\subsection{Representation-dependent tensor fields}

The tensor fields we have constructed may give rise to \emph{operator-valued} tensor fields if we replace the vector fields with corresponding elements in some operator representation of the Lie algebra Lie$(\G)$ of the group $\G$. In short: We may consider any Lie algebra representation $R$ of $\mbox{Lie}(\mathcal{G})\equiv T_e \G$ 
and replace $X_j$ with $R(X_j)$.\\
For instance, if $\mathcal{G}$ acts on a manifold $\mathcal{M}$, i.e.
\be \phi:\mathcal{G}\times\mathcal{M}\rightarrow \mathcal{M} \ee
we have a canonical action obtained from taking the tangent map
\be T\phi:T\mathcal{G}\times T\mathcal{M}\rightarrow T\mathcal{M}. \ee
We obtain in this way a Lie algebra homomorphism into the Lie algebra of vector fields on $\mathcal{M}$. Notice that the application of the tangent functor requires the action to be differentiable. If we replace the action of $\G$ on itself with the action on a generic manifold $\mathcal{M}$, we may formally write 
\be -\Phi(g)^{-1}d\Phi(g) \equiv R(X_j)\otimes\theta^j,\ee
the specific homomorphism depending on
\be \Phi: \mathcal{G} \rightarrow \text{Diff}(\mathcal{M})\ee
with the infinitesimal generators $R(X_j)$, 
\be R:\text{Lie}(\mathcal{G}) \rightarrow \text{Vect}(\mathcal{M}),\ee
from the Lie algebra of $\G$ into the module of vector fields on $\mathcal{M}$.\\
A corresponding tensor field on $\mathcal{G}$ is then provided by 
\be -\Phi(g)^{-1}d\Phi(g) \otimes \Phi(g)^{-1}d\Phi(g),\ee
yielding
\be (-R(X_j)\otimes R(X_k))\otimes \theta^j\otimes \theta^j.\ee 
To identify $R(X_j)\otimes R(X_k)$ with an `operator' we have several options. When the homomorphism maps to vector fields, the product $R(X_j)\otimes R(X_k)$ may be associated, for instance, either with a second order differential operator
\be L_{R(X_j)}(L_{R(X_k)}f)\ee
on a function $f \in \mathcal{F}(\mathcal{M})$
or with a bi-differential operator 
\be L_{R(X_j)}f_1(L_{R(X_k)}f_2)\ee
on pairs of functions $(f_1, f_2) \in \mathcal{F}(\mathcal{M})\times \mathcal{F}(\mathcal{M})$.\\    
Within the quantum setting, if we identify the manifold $\mathcal{M}$ say either in the Schr\"odinger picture as a Hilbert space $\H$ resp.\,the associated space of rays $\mathcal{R}(\H)$, or in the Heisenberg picture with the $\C^*$-algebra $\mathcal{\A}$ of dynamical variables, resp. the strata $D^k(\A)\subset \A^*$ of quantum states with fixed rank $k$, we may obtain several interesting tensor fields. Thus, if we consider
\be \mathcal{M}\equiv\begin{cases}
\H\\
 \mathcal{R}(\H)\\
\mathcal{A}\\
D^k(\A)\end{cases}
\ee
we may construct several tensor fields with values in the corresponding operator algebras.\\
In the following we will mainly focus on unitary representations
\be U: \mathcal{G}\rightarrow U(\H),\ee
as starting point. Here we will find the anti-Hermitian operator-valued left-invariant 1-form
\be-U(g)^{-1}dU(g)\equiv iR(X_j)\theta^j\ee
and construct a tensor field on $\mathcal{G}$
\be -U(g)^{-1}dU(g) \otimes U(g)^{-1}dU(g),\ee
yielding
\be R(X_j)R(X_k)\theta^j\otimes \theta^j,\ee 
where the operator $iR(X_j)$ is associated with the representation of the Lie algebra $\text{Lie}(\mathcal{G})$. This representation is equivalently defined by means of the representation of the enveloping algebra of the Lie algebra in the operator algebra $\A := End(\H)$. In particular, an element $X_j\otimes X_k$ in the enveloping algebra becomes then associated with a product 
\be R(X_j)R(X_k)\in \A:=End(\H).\ee
We may evaluate each of them by means of dual elements 
\be \rho\in \A^*,\ee
according to
\be \rho(R(X_j)R(X_k)) \equiv \Tr(\rho\, R(X_j)R(X_k))\in \C,\ee
yielding a complex-valued tensor field
\be \rho(R(X_j)R(X_k))\theta^j\otimes \theta^j\label{rho tensor}\ee
on the group manifold. The tensor field (\ref{rho tensor}) decompose into `classical' tensors in terms of a symmetric and an anti-symmetric part 
\be \rho([R(X_j),R(X_k)]_+)\theta^j\odot \theta^j\ee
\be \rho([R(X_j),R(X_k)]_-)\theta^j\wedge \theta^j.\ee
whose coefficients define real (resp.\,imaginary) scalar-valued functions 
\be \text{Lie}(\mathcal{G})\times \text{Lie}(\mathcal{G}) \times \A^*\rightarrow  \C\ee
which become bi-linear on the Lie algebra $\text{Lie}(\mathcal{G})$ once the dual element $\rho\in \A^*$ is fixed. With this map we also associate with two elements in $\text{Lie}(\mathcal{G})$, an element in $\A^*$.\\
A general operator-valued tensor field of order-k on a Lie group is defined by taking the k-th product of operator-valued left-invariant 1-forms
\be -U(g)^{-1}dU(g) \otimes U(g)^{-1}dU(g)\otimes...\otimes U(g)^{-1}dU(g),\ee
yielding
\be R(X_{j_1})R(X_{j_2})..R(X_{j_k})\theta^{j_1}\otimes \theta^{j_2}\otimes..\otimes \theta^{j_k}.\ee 
An evaluation of this higher rank operator-valued tensor field on a dual element $\rho\in\A^*$ according to
\be \rho(R(X_{j_1})R(X_{j_2})..R(X_{j_k}))\theta^{j_1}\otimes \theta^{j_2}\otimes..\otimes \theta^{j_k}\label{higher rank tensor}\ee 
provides a map 
\be \text{Lie}(\mathcal{G})\times \text{Lie}(\mathcal{G}) \times.. \times  \text{Lie}(\mathcal{G})\times \A^*\rightarrow  \C\ee
for each coefficient and therefore a corresponding linear functional-dependend multi-linear form on the Lie algebra.
This suggests the following definition:
\begin{Def}[Invariant operator-valued tensor field (IOVT)]\label{extrinsic tensors}
Let $\{\theta_j\}_{j\in J}$ be a basis of left-invariant 1-forms on $\mathcal{G}$, and let
$\{X_j\}_{j\in J}$ be a dual basis of left-invariant vector fields with $\{iR(X_j)\}_{j\in J}$, its representation in the Lie algebra $u(\H)$ of $U(\H)$ associated to a unitary representation $U:\mathcal{G}\rightarrow U(\H)$. The composition  
\be \bigg(\prod_{a=1}^k R(X_{j_a})\bigg) \bigotimes_{a=1}^k\theta^{j_a}\label{rank-k tensor}\ee 
defines then a covariant \emph{IOVT} of order-k on the Lie group $\G$, associating to any group element $g\in \G$ a map
\be T_g\G \times T_g\G.. \times T_g\G \times \A^*\rightarrow \C\ee
 and therefore a k-multilinear form on its Lie algebra $\text{Lie}(\mathcal{G})$ for any evaluation with a dual element $\rho\in \A^*$.
\end{Def}
\begin{Rem}
The evaluation of an IOVT becomes a \emph{quantum state-dependent tensor field}, once we restrict on dual elements $\rho\in \A^*$ which are positive and normalized. 
\end{Rem}
\begin{Rem}
This construction may be considered as generalization to Hermitian manifolds of Poincar\'e absolute and relative invariants within the framework of symplectic mechanics \cite{Arnold:76}.
\end{Rem}

\subsection{Non-linear actions and covariance matrix tensor fields}\label{Sums of IOVTS}

At this point we would like to relate the IOVT-construction to the pull-back tensor field 
\be \kappa_{\G}(\rho) :=(\rho(R(X_j)R(X_k))-  \rho(R(X_j))\rho(R(X_k))\theta^j\otimes \theta^k\label{ProjectivePBT}\ee 
which has been derived and applied in the previous part I of the work in \cite{Aniello:09} 
from a covariant tensor field
\be \frac{d\bar{z}^{j}\otimes dz^{j}}{\sum_j |z^{j}|^2}-\frac{z^{j}d\bar{z}^{j}\otimes \bar{z}^{k}dz^{k}}{(\sum_j |z^{j}|^2)^2}\label{ProjectiveHT}\ee
defined on a finite-dimensional punctured Hilbert space $\H_0\cong \C^{n+1}-\{0\}$. The IOVT coincides with the pull-back tensor construction whenever the dual element $\rho$ is restricted to be a \emph{pure} state.  While (\ref{ProjectiveHT}) is induced by the projection on the associated projective $\C P^{n}$ being endowed with the Fubini study metric, one finds that (\ref{ProjectivePBT}) is induced by a unitary action of the underlying Lie group $\G$ on a normalized fiducial vector $\ket{0}\in \H$ with $\rho\equiv \ket{0}\bra{0}$ \cite{Aniello:08, Aniello:09,Volkert:2010}.  The latter action is related to the co-adjoint action, infinitesimally defined by
\be R(X_j)\cdot \rho := [R(X_j), \rho].\ee
Formula (\ref{ProjectivePBT}) may be rewritten in terms of
\be \tilde{R}(X_j)\cdot \rho := [R(X_j)- \rho(R(X_j))\mbox{Id}_{\H}]\rho,\ee
so that $\kappa_{\G}(\rho)$ becomes
\be \rho(\tilde{R}(X_j)\tilde{R}(X_k))\theta^j\otimes \theta^k.\ee
This formula makes sense also when $\rho$ is not a pure state. The associated tensor coefficients $\rho(\tilde{R}(X_j)\tilde{R}(X_k))$ coincide with those of a covariance matrix, when evaluated on a linear functional $\rho \in \A^*$ which is positive and normalized. This follows directly from
$$  \rho(\tilde{R}(X_j)\tilde{R}(X_k)) =\rho\bigg([R(X_j)- \rho(R(X_j))\mbox{Id}_{\H}][R(X_k)- \rho(R(X_k))\mbox{Id}_{\H}]\bigg)$$
$$=\rho\bigg(R(X_j)R(X_k)+ \rho(R(X_j))\rho(R(X_k))\mbox{Id}_{\H}- \rho(R(X_j))R(X_k) - \rho(R(X_k))R(X_j)\bigg)$$
\be =\rho(R(X_j)R(X_k))-\rho(R(X_j))\rho(R(X_k)):=K_{jk}(\rho),\ee
where we used the normalization $\rho(\mbox{Id}_{\H})=1$.
It establishes an invariant \emph{covariance matrix tensor field}
\be  \kappa_{\G}(\rho)\equiv K_{jk}(\rho)\theta^j\otimes\theta^k,\ee
on the group manifold $\G$,  being decomposed into 
\be \kappa_{\G}= G+i\Omega,\ee
a symmetric part $G$, a part $L$, linear in $\rho$ and an anti-symmetric part $\Omega$ according to 
\be L(\rho) := \rho([R(X_j),R(X_k)]_+)\theta^j\odot \theta^k,\ee
\be G(\rho):= (L(\rho)-\rho(R(X_j)))\rho(R(X_k)))\theta^j\odot \theta^k,\label{G-CM}\ee
\be \Omega(\rho):= \rho([R(X_j),R(X_k)]_-)\theta^j\wedge \theta^k.\ee
The covariance matrix tensor field generalizes therefore the pull-back tensor field construction (\ref{ProjectivePBT}) on Lie groups applied to pure states, as considered in the previous papers
in \cite{Aniello:08, Aniello:09}, to (non-linear) operator-valued tensor fields on Lie groups
applied to mixed states.

\section{Evaluating IOVTs on composite systems}\label{Evaluating IOVTs on composite systems}

In the following we will discuss applications of IOVTs on mixed quantum states associated to n-level composite systems. For this purpose we recall some basic facts on the notion of a mixed state. \\
Within the Schr\"odinger picture we consider $\H_0=\H-\{0\}$ and the momentum map
\be \mu: \H_0 \rightarrow u^*(\H),\ee
\be \ket{\psi}\mapsto \ket{\psi}\bra{\psi}:=\rho_{\psi}\ee
on the real vector space $u^*(\H)$ of Hermitian operators.
This map has the image $\mu(\H_0)$,  given by the submanifold 
\be D^1(\H)\subset u^*(\H)\ee
of rank-1 Hermitian operators.  By restricting the momentum map to normalized vectors on the unit sphere $$ S_1(\H):=\{\ket{\p}\in \H|\sqrt{\braket{\p}{\p}}=1\}$$   of the Hilbert space, one finds an embedding $\iota$ of the space of rays $\mathcal{R}(\H)$ into $u^*(\H)$. The image 
\be \iota(\mathcal{R}(\H))=D_1^1(\H)\subset D^1(\H)\subset  u^*(\H)\label{pure states embedding}\ee 
becomes now the subset $D_1^1(\H)$ of rank-1 projection operators $\rho_{\psi_j}:=\rho_j$ and the convex combination
\be \rho = \sum_{j}p_j \rho_j,\,\,\,\,\,\sum_j p_j =1,\,\,\,\,\,p_j\in \R^+ \label{mixed qs def}\ee
yields the space of mixed quantum states which we denote here and in the following by $D(\H)$. Pure states are extremal states with respect to convex combinations. They may be characterized by the following properties when $\H=\C^n$.

\subsection{Pure, mixed and maximally entangled states}

Given a basis of generalized Pauli-matrices\footnote{These matrices are specified by being Hermitian and sharing the properties 
$$\sigma_0 = \mathds{1}_{n\times n},\quad
 \Tr(\sigma_j)= 0  \mbox{ for } j > 0$$  
$$ [\sigma_j,\sigma_k]_+ =\frac{2}{n}\delta_{jk}\sigma_0+d_{jkl} \sigma_l, \quad
[\sigma_j,\sigma_k]_- = c_{jkl}\sigma_l$$
where $c_{jkl}$ and $d_{jkl}$ denote full anti-symmetric and symmetric structure constants of the Lie-algebra $u(n)$ within the convention
\be [A,B]_+:=\2(AB+BA), \quad [A,B]_-:=\ii(AB-BA).\notag\ee
Note, that a trace-orthonormalization $\Tr(\sigma_j\sigma_k)= 2\delta_{jk}$ is implied in these properties by the decomposition
$$ \sigma_j\sigma_k = [\sigma_j,\sigma_k]_+ +i[\sigma_j,\sigma_k]_-.$$} $\sigma_j\in\{\sigma_j\}_{0\leq j\leq n^2-1}$  on the space of Hermitian matrices $u^*(n)$, we may expand any given mixed quantum state in the Bloch-representation
\be  \rho = \frac{1}{n}(\sigma_0 +m_j \s_j).\label{B-rep}\ee  
Here we find
\be \rho^2 =  \frac{1}{n^2}(\sigma_0 +2 m_k\s_k + m_j m_k \s_j \s_k)\ee
and therefore
\be \Tr(\rho^2)= \frac{1}{n^2}(n+2m_jm_k\delta_{jk}), \ee
where we used $\Tr(\s_j \s_k)=2\delta_{jk}$.
With $\rho^2=\rho$ for pure states and $\Tr(\rho)=1$ it follows:
\begin{Pro}\label{Purity-criteria}
For a given density state $\rho\in D(\C^n)$ in
the Bloch-representation (\ref{B-rep}), 
following statements are equivalent: 
\be \rho\text{ is pure, i.e. }\rho^2=\rho\tag{a}\ee
\be \sqrt{\sum_j m_j^2} = \sqrt{\frac{n(n-1)}{2}}.\tag{b}\label{Purity-criteria (b)}\ee  
\end{Pro} 
\begin{Rem}
Note that 
\be \sqrt{\sum_j m_j^2}:=||\rho||_2|_{su(n)} \ee
coincides with the Euclidean trace norm on the traceless part of the Bloch-representation (\ref{B-rep}).
\end{Rem}
On the other end,  `maximal mixed states' are defined to be the multiple of the identity
\be \rho^* := \frac{1}{n} \sigma_0.\ee
We find a first application of an invariant operator-valued tensor field (IOVT) considered in the previous section.
\begin{Pro}\label{Mixed-criteria}
Let 
\be \Omega := [R(X_j),R(X_k)]_-\theta^j\wedge \theta^k\ee
be an anti-symmetric \emph{IOVT} on the Lie group $U(n)$ in the defining representation. For a given density state $\rho\in D(\C^n)$ in the Bloch-representation (\ref{B-rep}), the following statements are equivalent: 
\be \rho\text{ is maximally mixed }\tag{a}\ee
\be \Omega(\rho)=0. \tag{b}\ee 
\be \sqrt{\sum_j m_j^2} = 0.\tag{c}\label{Mixed-criteria (c)}\ee
\end{Pro} 
\bp
In the defining representation $U(n)\times \C^n \rightarrow \C^n$ we have
\be \Omega = [\s_j,\s_k]_-\theta^j\wedge \theta^k= \sigma_l c_{jkl}\theta^j\wedge \theta^k.\ee  
For a given state $\rho\in D(\C^n)$ we find therefore 
\be  \Omega_{[jk]}(\rho) = \text{Tr}(\rho c_{jkl}\sigma_l).\ee
By  decomposing $\rho$ into Hermitian orthonormal matrices within the Bloch-representation 
\be \rho= \frac{1}{n}(\mathds{1}+\sum_{r=1}^{n^2-1} m_r\sigma_r),\label{red state}\ee
we find from the traceless property of the Hermitian matrices $c_{jkl}\sigma_l$ that
$$ \Omega_{[jk]}(\rho) = \text{Tr}(\rho  c_{jkl}\sigma_l)= \frac{1}{n}\text{Tr}((\mathds{1}+\sum_{r=1}^{n^2-1} m_r\sigma_r)  c_{jkl}\sigma_l)$$
$$ =\frac{1}{n}\text{Tr}(c_{jkl}\sigma_l+\sum_{r=1}^{n^2-1} c_{jkl}m_r \sigma_r\sigma_l) =\frac{1}{n}\text{Tr}(\sum_{r=1}^{n^2-1} c_{jkl}m_r \sigma_r\sigma_l)$$
\be=\frac{1}{n}\sum_{r=1}^{n^2-1} c_{jkl}m_r\text{Tr}(\sigma_r\sigma_l)=\frac{2}{n}\sum_{r=1}^{n^2-1} c_{jkl}m_r \delta_{rl}= \frac{2}{n}\sum_{l=1}^{n^2-1} c_{jkl} m_l.\ee
With
\be \sum_{l=1}^{n^2-1} c_{jkl} m_l\sigma_k =\sum_{l=1}^{n^2-1}[\sigma_j, m_l\sigma_l]\ee
we find that 
\be \sum_{l=1}^{n^2-1} c_{jkl} m_l=0\ee
iff 
\be \sum_{l=1}^{n^2-1} c_{jkl} m_l\sigma_k=0.\label{condition}\ee
Since the Lie algebra of $SU(n)$ is perfect, i.e.
\be [su(n),su(n)]=su(n),\ee
it follows that the condition (\ref{condition}) holds true if and only if 
\be m_l = 0 \text{ for all }l.\ee
This implies $\Omega_{[jk]}(\rho)=0$ and, according to (\ref{red state}), a state 
\be \rho=\frac{1}{n}\mathds{1}\ee 
from which one concludes the statement.
\ep
\begin{Cor}\label{Cor L ent = max ent}
Lagrangian submanifolds of pure bi-partite quantum states with same Schmidt coefficients define  maximally entangled pure states.
\end{Cor}
\bp
Since Schmidt-equivalence classes with equal distributed Schmidt-coefficients are maximal entangled according to the von Neumann entropy measure, we can directly conclude the statement by applying Proposition \ref{Mixed-criteria} on reduced density states within Proposition 1 (a) in \cite{Aniello:09}.\ep
We remark that this is a stronger statement than the one made by Bengtsson \cite{Bengtsson:2007}, where it has been concluded from dimensional arguments that maximal entangled states provide a Lagrangian submanifold. We have shown here that there are no further Lagrangian submanifolds, which provide Schmidt equivalence classes of entangled states being less entangled than maximal.

\subsection{Entanglement monotone candidates from IOVTs on $U(n)\times U(n)$}

According to Vidal \cite{Vidal:2000}, one defines entanglement monotones as functions
\be f:D(\H_A\otimes \H_B) \rightarrow \R_+ \ee 
which do not increase under the set LOCC of local quantum operations being assisted by classical communication. A necessary (although not sufficient) way to identify entanglement monotones is therefore given by functions on $D(\H_A\otimes \H_B)$ being invariant under the local unitary group of transformations $U(\H_A)\times U(\H_B)$. This suggests to define \emph{entanglement monotone candidates} by functions on $D(\H_A\otimes \H_B)$ satisfying
\be f(U\rho U^{\dagger}) = f(\rho) \mbox{ for all } U\in U(\H_A)\times U(\H_B). \ee
In this necessary strength, we propose in the following entanglement monotones candidates by identifying constant functions on local unitary orbits of entangled quantum states, arising from invariant IOVTs constructed on $U(\H_A)\times U(\H_B)$. To map the latter invariant IOVTs to invariant functions, we use an inner product defined on the space of `classical' quantum state evaluated IOVTs of fixed order as follows. 
Based on definition \ref{extrinsic tensors} we consider the representation R-dependent IVOT of order $k$
\be \theta_R := \bigg(\prod_{a=1}^k R(X_{i_a})\bigg) \bigotimes_{a=1}^k\theta^{i_a}\ee 
on a Lie group $\G$. After evaluating it with a state 
\be \theta_R \mapsto \rho(\theta_R):=\theta^{\rho}_R =\rho\bigg(\prod_{a=1}^k R(X_{i_a})\bigg) \bigotimes_{a=1}^k\theta^{i_a}\ee
we may consider for compact Lie groups the Hermitian inner product
\be \braket{\theta^{\rho}_R}{\theta^{\rho}_R}:= \rho\bigg(\prod_{a=1}^k R(X_{i_a})\bigg)^*\rho\bigg(\prod_{a=1}^k R(X_{j_a})\bigg) \braket{\bigotimes_{a=1}^k\theta^{i_a}}{\bigotimes_{a=1}^k\theta^{j_a}} \ee
By defining 
\be  \rho\bigg(\prod_{a=1}^k R(X_{i_a})\bigg):=T^{\rho}_{i_1...i_k}\in \C\ee
and using
\be  \braket{\bigotimes_{a=1}^k\theta^{i_a}}{\bigotimes_{a=1}^k\theta^{j_a}}=\prod_{a=1}^k \braket{\theta^{i_a}}{\theta^{j_a}} =\prod_{a=1}^k \delta_{i_a j_a}\ee 
one finds 
\be \braket{\theta^{\rho}_R}{\theta^{\rho}_R}=\sum_{i_1,...,i_k} |T^{\rho}_{i_1...i_k}|^2\in \R,\ee
a quadratic function on tensor coefficients.\\
By considering this $\G$-invariant construction in particular on the local unitary group $\G\equiv U(\H_A)\times U(\H_B)$ for a given representation $R$, we may introduce a $R$-\emph{class} of entanglement monotone candidates by the set of polynomial functions 
\be f^R_k(\rho):=\sum_n a_n  \braket{\theta^{\rho}_R}{\theta^{\rho}_R}^n,\quad a_n\in \R\ee
made out of these quantum state dependent Hermitian products for all IOVTs of fixed order $k\in \mathbb{N}$. 
This class can be enlarged in different directions, either by considering polynomials involving IOVTs with variable order $k$ or by considering alternative realizations $\tilde{R}$, going possibly also beyond unitary representations.

\begin{Rem}
This framework may be generalized to multi-partite quantum systems by using representations of corresponding higher order products of unitary groups, i.e.\,by taking into account IOVTs on $\G\equiv U(\H)^{\times r}$ with $r\ge 2$. 
\end{Rem} 

\subsubsection{Relation to known entanglement monotones $n=2$}\label{rel to known em}

To compare these functions with possibly known entanglement monotones from the literature, we restrict the following discussion on the case of two qubits.
To review here some of the well established entanglement-monotones, consider first a real parametrization  of the convex subset of bi-partite states
\be \rho \equiv \frac{1}{4}(\s_0\otimes\s_0 +  n_j \s_j\otimes\s_0 +m_k \s_0\otimes\s_k +C_{jk} \s_j\otimes\s_k).\label{Fano-Form1states}\ee
with 
\be n_j :=\Tr(\rho^A\sigma_j) = \Tr(\rho \sigma_j\otimes \mathds{1}), \qquad m_k :=\Tr(\rho^B\sigma_k) = \Tr(\rho \mathds{1}\otimes \sigma_k) \ee
\be C_{jk} :=\Tr(\rho \sigma_j\otimes\sigma_k). \ee
By considering the spin-flip operation 
\be\widetilde{\rho}:= \sigma_2\otimes\sigma_2 \rho^* \sigma_2\otimes\sigma_2.\label{spin flip}\ee
one finds
\be  4\Tr(\rho\widetilde{\rho})=1-\sum_{j=1}^3 m_j^2-\sum_{j=1}^3 n_j^2 +\sum_{j,k =1}^3 C_{j,k}^2,\label{SL-inv} \ee
which turns out to be a central quantity for entanglement quantification as follows: For mixed states $\rho\in D(2)$ one finds that the singular values of $\rho\widetilde{\rho}$ yield monotones for constructing the \emph{concurrence}
\be C(\rho):=\mbox{max}[\lambda_4-\lambda_3-\lambda_2-\lambda_1, 0], \quad\lambda_j\in\mbox{Spec}(\rho\widetilde{\rho}), \quad \lambda_4\le..\le \lambda_1,\label{concu}\ee  providing the \emph{entanglement measure of formation} \cite{Wootters:1997id,Coffman:1999jd, Jaeger:2003}. Moreover, one recovers the purity measure
\be  \Tr(\rho^2)=\Tr(\rho\widetilde{\rho}),\quad \mbox{ if } \widetilde{\rho}=\rho,\ee
i.e., if the given state is invariant under the spin-flip operation (\ref{spin flip}). In this case it provides a \textit{complementary} entanglement monotone to the linear entropy 
\be 1-\Tr(\rho^2),\ee
as underlined in \cite{Jaeger:2003}. As the name suggests,  the linear entropy may be considered to be an approximation to the von Neumann entropy
\be -\Tr(\rho \ln\rho),\ee
for $\ln \rho \approx \rho-1$.
For more details on entanglement monotones we refer to the seminal paper of Vidal \cite{Vidal:2000}, resp.\,\cite{Bengtsson:2006} and references therein.\\
In the following we would like to show how some of these  entanglement monotones may be recovered, up to additional terms, from an evaluation of (\ref{Fano-Form1states}) on the order-2 IOVT
\be \theta_R :=  R(X_j)R(X_k)\theta^j\odot \theta^k\ee
being decomposed into a symmetric 
\be G_R:= [R(X_j), R(X_k)]_+\theta^j\odot \theta^k\ee
and an anti-symmetric IOVT
\be \Omega_R:= [R(X_j), R(X_k)]_-\theta^j\wedge \theta^k\ee
on $\mathcal{G}=SU(2)\times SU(2)$ associated to a product representation $U:\mathcal{G}\rightarrow U(4)$ on $\C^2\otimes \C^2$ with
\be R(X_j) =
 \begin{cases}
 \sigma_j\otimes \mathds{1} &\text{ for } 0 \le j \le 3\\
 \mathds{1}\otimes  \sigma_{j-4} &\text{ for } 4 \le j \le 7. 
\end{cases}\label{U(2)h-basis}
 \ee
Computing the inner product of the tensor field
\be \theta^{\rho}_R :=G^{\rho}_R+\Omega^{\rho}_R,\ee
as proposed in the previous section on $\theta^{\rho}_R$, separately on the symmetric and the anti-symmetric part, we find\footnote{The quantum state dependent coefficients may be subsumed for the symmetric part
\be
(G^{\rho}_{(jk)})_R=\left(
\begin{array}{llllll}
 1 & 0 & 0 & C_{1,1} & C_{1,2} & C_{1,3} \\
 0 & 1 & 0 & C_{2,1} & C_{2,2} & C_{2,3} \\
 0 & 0 & 1 & C_{3,1} & C_{3,2} & C_{3,3} \\
 C_{1,1} & C_{2,1} & C_{3,1} & 1 & 0 & 0 \\
 C_{1,2} & C_{2,2} & C_{3,2} & 0 & 1 & 0 \\
 C_{1,3} & C_{2,3} & C_{3,3} & 0 & 0 & 1
\end{array}
\right)
\ee
whereas the anti-symmetric part yields
\be (\Omega^{\rho}_{[jk]})_R=\left(
\begin{array}{llllll}
 0 & m_3 & -m_2 & 0 & 0 & 0 \\
 -m_3 & 0 & m_1 & 0 & 0 & 0 \\
 m_2 & -m_1 & 0 & 0 & 0 & 0 \\
 0 & 0 & 0 & 0 & n_3 & -n_2 \\
 0 & 0 & 0 & -n_3 & 0 & n_1 \\
 0 & 0 & 0 & n_2 & -n_1 & 0
\end{array}
\right).\ee
}
\be \frac{1}{8}\big(\braket{G^{\rho}_R}{G^{\rho}_R}+ (-1)^j\braket{\Omega^{\rho}_R}{\Omega^{\rho}_R}\big) -\frac{1}{2}= \begin{cases}
 \mbox{Tr}(\rho^2) &\text{ for } j=0\\
 \mbox{Tr}(\rho\tilde{\rho}) &\text{ for } j=1. 
\end{cases}\label{tangle measure}\ee
Hence, by setting 
\be f_2^R(\rho)\equiv \braket{\theta^{\rho}_R}{\theta^{\rho}_R} \ee
we may identify
\be \frac{3}{2}- \frac{1}{8}f_2^R(\rho) = 1-
 \mbox{Tr}(\rho^2) \ee
as an approximation to the von Neumann entropy. 

\subsubsection{Alternative entanglement monotone candidates for $n=2$}\label{new m}
To get into account a possible approximation to the concurrence (\ref{concu}) we shall consider an IOVT build by a non-linear realization, 
\be \tilde{R}(X_j)\cdot \rho := [R(X_j)- \rho(R(X_j))\mathds{1}]\rho.\ee
Here we find 
$$f_2^{ \tilde{R}}(\rho) =6+\sum_{j=1}^3 m_j^4+\sum_{j=1}^3 n_j^4 +2\sum_{j,k =1}^3 C_{j,k}^2$$
$$ +2m_1^2 \left(m_2^2+m_3^2+n_1^2+n_2^2+n_3^2\right) +2 m_3^2 \left(n_1^2+n_2^2+n_3^2\right)+2 m_2^2
   \left(m_3^2+n_1^2+n_2^2+n_3^2\right)$$
\be -4 m_1 \left(n_1 C_{1,1}+n_2 C_{1,2}+n_3 C_{1,3}\right)$$ $$-4 m_2 \left(n_1 C_{2,1}+n_2 C_{2,2}+n_3 C_{2,3}\right)-4 m_3 \left(n_1
   C_{3,1}+n_2 C_{3,2}+n_3 C_{3,3}\right).\ee   
While the anti-symmetric IOVT remains unchanged for this non-linear realization, one finds here for the the inner product on the symmetric tensor field\footnote{The coefficients read here $$(G^{\rho}_{[jk]})_{\tilde{R}} =\left(
\begin{array}{llllll}
 1-m_1^2 & -m_1 m_2 & -m_1 m_3 & C_{1,1}-m_1 n_1 & C_{1,2}-m_1 n_2 & C_{1,3}-m_1 n_3 \\
 -m_1 m_2 & 1-m_2^2 & -m_2 m_3 & C_{2,1}-m_2 n_1 & C_{2,2}-m_2 n_2 & C_{2,3}-m_2 n_3 \\
 -m_1 m_3 & -m_2 m_3 & 1-m_3^2 & C_{3,1}-m_3 n_1 & C_{3,2}-m_3 n_2 & C_{3,3}-m_3 n_3 \\
 C_{1,1}-m_1 n_1 & C_{2,1}-m_2 n_1 & C_{3,1}-m_3 n_1 & 1-n_1^2 & -n_1 n_2 & -n_1 n_3 \\
 C_{1,2}-m_1 n_2 & C_{2,2}-m_2 n_2 & C_{3,2}-m_3 n_2 & -n_1 n_2 & 1-n_2^2 & -n_2 n_3 \\
 C_{1,3}-m_1 n_3 & C_{2,3}-m_2 n_3 & C_{3,3}-m_3 n_3 & -n_1 n_3 & -n_2 n_3 & 1-n_3^2
\end{array}
\right).$$}
$$ \braket{G^{\rho}_{\tilde{R}}}{G^{\rho}_{\tilde{R}}}=\left(m_1^2-1\right){}^2+2 m_1^2 m_2^2+\left(m_2^2-1\right){}^2+2 m_1^2 m_3^2+2 m_2^2
   m_3^2$$ $$+\left(m_3^2-1\right){}^2+\left(n_1^2-1\right){}^2+2 n_1^2 n_2^2+\left(n_2^2-1\right){}^2+2 n_1^2 n_3^2+2 n_2^2
   n_3^2$$ $$+\left(n_3^2-1\right){}^2+2 \left(C_{1,1}-m_1 n_1\right){}^2+2 \left(C_{1,2}-m_1 n_2\right){}^2$$ $$+2 \left(C_{1,3}-m_1
   n_3\right){}^2+2 \left(C_{2,1}-m_2 n_1\right){}^2+2 \left(C_{2,2}-m_2 n_2\right){}^2$$ $$+2 \left(C_{2,3}-m_2 n_3\right){}^2+2
   \left(C_{3,1}-m_3 n_1\right){}^2+2 \left(C_{3,2}-m_3 n_2\right){}^2+2 \left(C_{3,3}-m_3 n_3\right){}^2. $$
In this way, we recover by means of those terms involving
\be \sum_{j,k} C_{j,k}-m_j n_k\ee
an entanglement monotone candidate which has been considered and applied on pure states within part I of the underlying work in (3.37), (3.38), p.10 in \cite{Aniello:09}.\\
In section \ref{Werner quantitative} we will outline an application of these invariant functions for a \emph{quantitative} description of entanglement associated to  subclass of two qubit states. Before coming to this point we will establish in general terms a link between a class of computable separability criteria, being useful for a \emph{qualitative} description of entanglement,  and these type of operator-valued tensor fields defined on the Lie group $U(n)\times U(n)$ for arbitrary finite-dimensional bi-partite systems.

\subsection{Separability criteria from IOVTs on $U(n)\times U(n)$}\label{deVicente-criteria}

Let us restart here by recalling the notion of separability for mixed quantum states (see e.g. \cite{Gurvits:2003}\cite{Bengtsson:2006}). By considering a product Hilbert space $\H\equiv \H_A\otimes\H_B$ we find a convex set of mixed states $D(\H_A\otimes\H_B)$. An arbitrary element will have the form
\be \rho = \sum_{j}p_j \rho_j,\,\,\,\,\,\sum_j p_j =1,\,\,\,\,\,p_j\in \R^+,\ee
with 
\be \rho_j \in D_1^1(\H_A\otimes\H_B)\cong \mathcal{R}(\H_A\otimes\H_B).\ee
Each rank-1 projector  is called \emph{separable} if it can be written as
\be \rho_j = \rho^A_j\otimes \rho^B_j\ee
with $\rho^s_j \in D^1_1(\H_s), s\in \{A,B\}$. For density states we have:

\begin{Def}[Separable and entangled mixed states]
A mixed state $\rho\in D(\H_A\otimes\H_B)$ is called separable if 
\be \rho = \sum_{j=1}p_j \rho^A_j\otimes \rho^B_j,\,\,\,\,\,\sum_j p_j =1,\,\,\,\,\,p\in \R^+ \label{ent def}\ee
with $\rho^s_j \in D^1_1(\H_s), s\in \{A,B\}$, otherwise entangled.
\end{Def}
An approach to mixed states separability criteria is known to be given by means of the Bloch-representation  \cite{deVicente:2007}. As starting point for deriving these criteria one finds: 
\begin{Pro}\label{Not FAPP-criteria mixed}
A given state $ \rho \in D(\H_A\otimes \H_B)\cong D(\C^n\otimes \C^n)$ in the Bloch-representation
\be \rho \equiv \frac{1}{n^2}(\s_0\otimes\s_0 +  n_j \s_j\otimes\s_0 +m_k \s_0\otimes\s_k +C_{jk} \s_j\otimes\s_k).\label{Fano-Form1}\ee
with 
\be n_j :=\Tr(\rho^A\sigma_j) = \Tr(\rho \sigma_j\otimes \mathds{1}), \qquad m_k :=\Tr(\rho^B\sigma_k) = \Tr(\rho \mathds{1}\otimes \sigma_k) \ee
\be C_{jk} :=\Tr(\rho \sigma_j\otimes\sigma_k). \ee
is separable iff there exists a decomposition into
\be n_j = \sum_{i}p_i n_j^i, \qquad m_k = \sum_{i}p_i m_k^i,  \qquad C_{jk} = \sum_{i}p_i n_j^i m_k^i \label{Not FAPP-criteria mixed t}\ee
with Bloch-representation coefficients  
\be n_j^i := \Tr(\rho_i^A\sigma_j), \qquad m_k^i :=\Tr(\rho_i^B\sigma_k).\ee
associated to pure states $\rho^s_i \in D^1_1(\H_s), s\in \{A,B\}$.
\end{Pro} 
\bp
By considering the pure states
\be  \rho_i^A = \frac{1}{n}(\sigma_0 +n_j^i \s_j),\qquad  \rho_i^B = \frac{1}{n}(\sigma_0 +m_k^i \s_k)\ee
within a separable mixed state
\be \rho =\sum_i p_i \rho_i^A \otimes \rho_i^B \ee
one finds
$$ \rho = \frac{1}{n^2}\sum_i p_i (\sigma_0 +n_j^i \s_j)\otimes (\sigma_0 +m_k^i \s_k) $$
$$ = \frac{1}{n^2}\sum_i p_i(\s_0\otimes\s_0 +  n_j^i \s_j\otimes\s_0 +m_k^i \s_0\otimes\s_k + n_j^i m_k^i \s_j\otimes\s_k) $$
$$ = \frac{1}{n^2}\bigg(\sum_i p_i\s_0\otimes\s_0 +  \sum_i p_i n_j^i \s_j\otimes\s_0 +\sum_i p_i m_k^i \s_0\otimes\s_k +\sum_i p_i n_j^i m_k^i \s_j\otimes\s_k\bigg).$$
By comparing the latter expression with a given state $\rho \in D(\H_A\otimes \H_B)$ in the Bloch-representation (\ref{Fano-Form1}) one concludes the statement.
\ep
It has been shown that this proposition implies several either sufficient or necessary separability criteria of computable nature \cite{deVicente:2007}. Without proof, we restate here one of the \emph{sufficient} criteria:

\begin{Pro}\label{Sufficient de Vicente}
A state $ \rho \in D(\C^n\otimes \C^n)$ in the Bloch-representation 
\be \rho \equiv \frac{1}{n^2}(\s_0\otimes\s_0 +  n_j \s_j\otimes\s_0 +m_k \s_0\otimes\s_k +C_{jk} \s_j\otimes\s_k)\ee
is separable if it fulfills the inequality
\be  \sqrt{\frac{2(n-1)}{n}}\Bigg(\sqrt{\sum_j n_j^2}+\sqrt{\sum_k m_k^2}\Bigg) + \frac{2(n-1)}{n}\Tr(\sqrt{C^{\dagger}C}) \le 1\ee 
with a coefficient matrix $C:= (C_{jk})_{j,k \in J}$.
\end{Pro}
In this regard we find a class of computable separability criteria from IOVTs, which are directly linked with the \emph{necessary} criteria in the Bloch-representation as proposed by de Vicente \cite{deVicente:2007}. Our translation in geometric terms reads\footnote{For details on the proof see \cite{Volkert:2010, deVicente:2007}.}:

\begin{The}\label{de Vicente 1}
Let 
\be  L:= [R(X_j), R(X_k)]_+\theta^j\odot \theta^k\ee
be a symmetric \emph{IOVT} on $\mathcal{G}=U(n)\times U(n)$ associated to a tensor product representation $U:\mathcal{G}\rightarrow U(n^2)$.
The evaluation of the coefficients
\be L_{(jk)}(\rho):= \Tr(\rho [R(X_j), R(X_k)]_+ ),\label{L}\ee  
on a state $\rho \in D(\C^n\otimes\C^n)$
implies then for  \be L_{(jk)}|_{j,k\in J}(\rho):= C_{jk}\ee in  $J:=\{j,k|1 \le j \le n^2-1; n^2 \le k \le 2n^2-2\}$ the inequality 
\be \|C\|_{KF}:=\Tr(\sqrt{C^{\dagger}C}) \le \frac{n(n-1)}{2} \label{ineq}\ee 
if $\rho$ is separable. 
\end{The}
To identify the geometric structure which prevents this necessary criterion from being also a sufficient criterion, one may focus on the \emph{anti-symmetric} rather then the symmetric IOVT within the following criterion: 

\begin{The}\label{de Vicente 2}
Let 
\be  \Omega:= [R(X_j), R(X_k)]_-\theta^j\wedge\theta^k\ee
be an anti-symmetric \emph{IOVT} on $\mathcal{G}=U(n)\times U(n)$ associated to a product representation $U:\mathcal{G}\rightarrow U(n^2)$.
A state $\rho \in D(\C^n\otimes\C^n)$ with
\be \Omega(\rho)=0, \ee 
is separable if it fulfills the inequality
\be  \frac{2(n-1)}{n}\|C\|_{KF}\le 1,\ee 
with a coefficient matrix $C$, defined as in Theorem \ref{de Vicente 1} by the evaluation on a symmetric \emph{IOVT}.
\end{The}
\bp
According to the product representation\footnote{See also section 3.1 in \cite{Aniello:09}.} one finds that the non-trivial coefficients are given in the block elements $\Omega_{[jk]}|_{j,k\in I}:= \Omega^{s}_{[jk]}$ with $s\in \{A, B\}$ of the coefficient matrix $(\Omega_{[jk]})$, defined by
\be \Omega^{A}_{[jk]}(\rho)=   \Tr(\rho[\sigma_j,\sigma_k]_-\otimes \mathds{1}),\ee
\be \Omega^{B}_{[jk]}(\rho)=   \Tr(\rho\mathds{1}\otimes [\sigma_{j-n^2},\sigma_{k-n^2}]_-).\ee
Both cases read
\be \Omega^{s}_{[jk]}= 
\text{Tr}(\rho^s  c_{jkl}\sigma_l),\label{S-mix=02}\ee
with the partial traces resp.\,the reduced density matrices 
\be \rho^s\equiv \text{Tr}_s(\rho),\,\,\, s\in \{A,B\}.\ee
By applying Proposition \ref{Mixed-criteria} in Proposition \ref{Sufficient de Vicente} it follows the statement.
\ep
For bi-partite 2-level systems the inequalities in the last two theorems coincide:
 
\begin{Cor}\label{de Vicente 1 Cor}
The inequality 
\be \|C\|_{KF} \le 1\ee
becomes both a sufficient and necessary separability criterion for the case $n=2$ with maximally mixed subsystems \cite{deVicente:2007}. 
\end{Cor}

\section{Applications on subclasses of states in $n=2$}\label{Applications on subclasses of states in n=2}

\subsection{A one-parameter subclass of entangled states: Werner states}

Let us apply the last two subsections on an explicit example. Consider for this purpose a convex combination of a rank-1 projector\footnote{Also known as Bell state, i.e.\,a maximal entangled pure state associated to the vector
$$ \ket{\phi^+}:=\frac{1}{\sqrt{2}}\begin{pmatrix}
1 \\
0 
\end{pmatrix}\otimes \begin{pmatrix}
1 \\
0
\end{pmatrix} + \frac{1}{\sqrt{2}}\begin{pmatrix}
0 \\
1 
\end{pmatrix}\otimes \begin{pmatrix}
0 \\
1
\end{pmatrix}\in \C^2\otimes \C^2.$$} $\ket{\phi^+}\bra{\phi^+}$ with
a maximal mixed state \be\rho^*:=\frac{1}{4}\mathds{1},\ee
according to  
\be \rho_W:=x\ket{\phi^+}\bra{\phi^+}+(1-x)\rho^*\ee
with $x\in [0,1]$, i.e.\,a density state $\rho_W\in D(\C^2\otimes \C^2)$, which in the literature is referred as the class of Werner states \cite{Werner:1989zz}.

\subsubsection{Qualitative description}

By evaluating 
\be  \rho_W \in D(\C^2\otimes \C^2),\ee
on a symmetric IOVT
\be  L:=[R(X_j), R(X_k)]_+\theta^j\odot \theta^k\ee
associated to a product representation $SU(2)\times SU(2)\rightarrow U(4)$ we find a matrix of coefficients 
\be (L_{(jk)})(\rho_W)= \left(
\begin{array}{llllll}
 1 & 0 & 0 & x & 0 & 0 \\
 0 & 1 & 0 & 0 & -x & 0 \\
 0 & 0 & 1 & 0 & 0 & x \\
 x & 0 & 0 & 1 & 0 & 0 \\
 0 & -x & 0 & 0 & 1 & 0 \\
 0 & 0 & x & 0 & 0 & 1
\end{array}
\right), \label{L on rhoW}\ee
defined on the 6-dimensional real Lie algebra of $SU(2)\times SU(2)$. A decomposition 
\be (L_{(jk)}):=\left(\begin{array}{cc}A & C \\C & B\end{array}\right),\ee
implies
\be C = \left(
\begin{array}{lll}
 x & 0 & 0 \\
 0 & -x & 0 \\
 0 & 0 & x
\end{array}
\right). \label{L on rhoW2}\ee  
The latter is identical to the tensor coefficients $L_{(jk)}$ for $1 \le j \le 3$ and
 $5 \le k \le 6$.  By computing the Ky Fan Norm of $C$ one finds
\be \text{Tr}(\sqrt{C^{\dagger}C}) = 3 x,\ee
thus we conclude according to Corollary \ref{de Vicente 1 Cor} that $\rho_W$ is separable iff
\be x \le \frac{1}{3},\label{1/3}\ee
since the evaluation of $\rho_W$ on an anti-symmetric IOVT
\be  \Omega:=[R(X_j), R(X_k)]_+\theta^j\wedge \theta^k\ee
associated to a product representation $SU(2)\times SU(2)\rightarrow U(4)$ yields
\be \Omega(\rho_W)=0.\ee
The latter condition can be checked by using the fact that a convex combination of states gives rise to convex combination of corresponding anti-symmetric tensors  
\be \Omega(\rho_W):=x \Omega(\ket{\phi^+}\bra{\phi^+})+(1-x) \Omega(\rho^*),\ee
which is zero in each term, i.e.\,for the maximal entangled pure state $\ket{\phi^+}\bra{\phi^+}$ and for the  maximal mixed state $\rho^*$, according to Proposition \ref{Mixed-criteria} and Corollary \ref{Cor L ent = max ent}.\\
Similarly, we may perform also on the symmetric tensor $L(\rho_W)$
a decomposition into a convex sum of symmetric tensors
\be  L(\rho_W)=  x L(\ket{\phi^+}\bra{\phi^+})+(1-x) L(\rho^*).\ee
As a curious fact we observe in this regard the following: If we invert the sign of the first term associated to the pure state $\ket{\phi^+}\bra{\phi^+}$ we may define a distinguished symmetric tensor construction
\be  \widetilde{L}(\rho_W):= -x L(\ket{\phi^+}\bra{\phi^+})+(1-x) L(\rho^*)\ee
yielding the coefficient matrix
\be
 (\widetilde{L}_{(jk)}(\rho_W))=\left(
\begin{array}{llllll}
 1-2 x & 0 & 0 & -x & 0 & 0 \\
 0 & 1-2 x & 0 & 0 & x & 0 \\
 0 & 0 & 1-2 x & 0 & 0 & -x \\
 -x & 0 & 0 & 1-2 x & 0 & 0 \\
 0 & x & 0 & 0 & 1-2 x & 0 \\
 0 & 0 & -x & 0 & 0 & 1-2 x
\end{array}
\right).\ee
A diagoanlization of this matrix reads
\be (\widetilde{L}'_{(jk)}(\rho_W))=\left(
\begin{array}{llllll}
 1-3 x & 0 & 0 & 0 & 0 & 0 \\
 0 & 1-3 x & 0 & 0 & 0 & 0 \\
 0 & 0 & 1-3 x & 0 & 0 & 0 \\
 0 & 0 & 0 & 1-x & 0 & 0 \\
 0 & 0 & 0 & 0 & 1-x & 0 \\
 0 & 0 & 0 & 0 & 0 & 1-x
\end{array}
\right)\ee
which provides a direct characterization of entanglement according to (\ref{1/3}), where we conclude that  $\rho_W$ is separable iff the symmetric tensor $\widetilde{L}$ has positive definite signature.
 
\subsubsection{Quantitative description}\label{Werner quantitative}

\begin{figure}[htp]
\centerline{\includegraphics[scale=0.70]{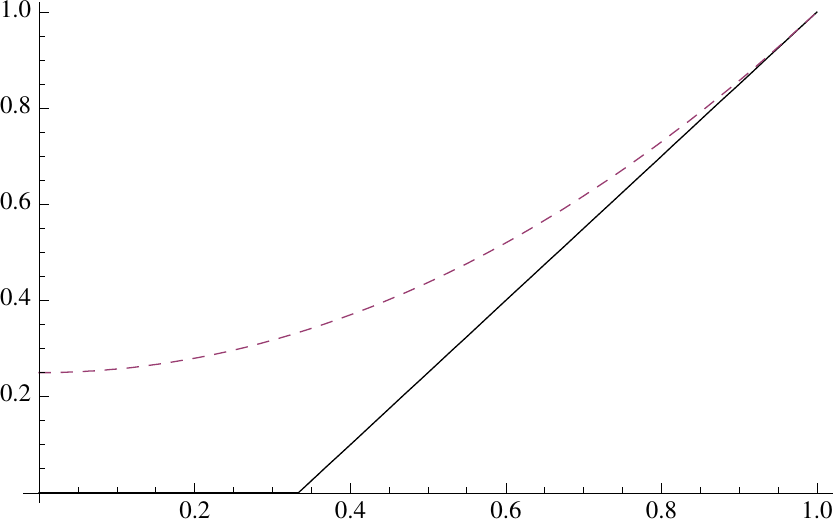}}
\vspace*{8pt}
\caption{Separable states take zero values, while entangled states take positive values according to the concurrence $\mathcal{C}(\rho_W)$ (straight line). The purity $\Tr(\rho_W\widetilde{\rho_W})=\Tr(\rho_W^2)$ yields an entanglement monotone, which provides a fair approximation of the concurrence $\mathcal{C}(\rho_W)$ for $x\rightarrow 1$.  \label{fig1}}
\end{figure}

By considering the spin-flip operation 
\be\widetilde{\rho}:= \sigma_2\otimes\sigma_2 \rho^* \sigma_2\otimes\sigma_2,\label{spin flip2}\ee
and the square root eigenvalues $\lambda_{j}$ of 
\be \rho\widetilde{\rho},\label{pp'}\ee 
one finds a quantitative description of mixed state bi-partite entanglement for two qubits in terms of the \emph{concurrence} \cite{Wootters:1997id,Coffman:1999jd}, defined by 
\be \mathcal{C}(\rho):= \mbox{max}(0, \lambda_{1}-\lambda_{2}-\lambda_{3}-\lambda_{4}),\ee
with $\lambda_{j}>\lambda_{j+1}$.
Due to the invariance of Werner states 
\be \widetilde{\rho_W}=\rho_W\ee
under spin-flip transformation, it may instructive to take into account the relation to the values of the purity $\Tr(\rho^2)$, which has been recovered here in section \ref{rel to known em} by the function 
\be  f_2^R(\rho)\equiv \braket{\theta^{\rho}_R}{\theta^{\rho}_R} \ee
with $\theta^{\rho}_R=\rho(R(X_j)R(X_k)\theta^j\otimes \theta^k$. According to (\ref{tangle measure}) we find 
$$f_2^{ R}(\rho_W) = 8\Tr(\rho_W^2)+4= 6 \left(x^2+1\right).$$
To compare this function with entanglement monotones related to the concurrence we find 
\be \mathcal{C}(\rho_W)= \mbox{max}\bigg(0, \frac{1}{2} (3 x-1)\bigg).\ee
This function is illustrated together with the purity in figure \ref{fig1}.

\subsection{A two-parameter subclass of entangled states}

\begin{figure}[htp]
\centerline{\includegraphics[scale=0.70]{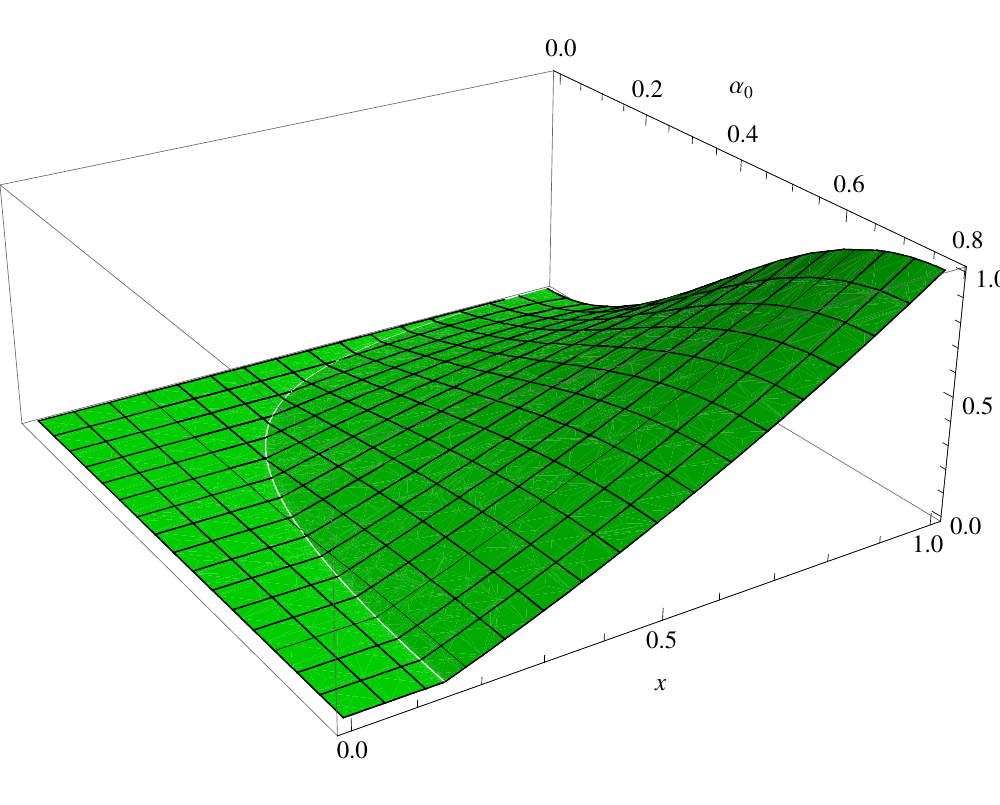}}
\vspace*{8pt}
\caption{Separable states take zero values, while entangled states take positive values according to the concurrence $\mathcal{C}(\rho_{x, \alpha_0})$. \label{fig2}}
\end{figure}

\begin{figure}[htp]
\centerline{\includegraphics[scale=0.70]{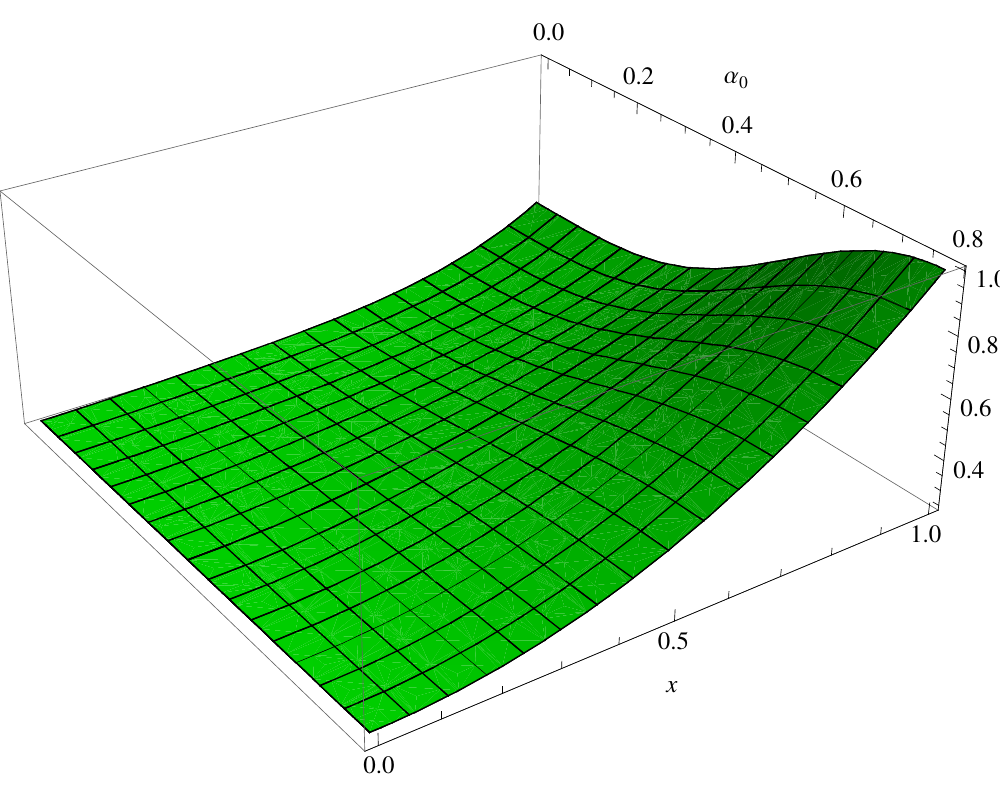}}
\vspace*{8pt}
\caption{A possible approximation of the concurrence measure derived from an non-linear realization-dependent IOVT, giving rise to the invariant $\mathcal{D}(\rho_{x, \alpha_0})$. \label{fig3}}
\end{figure}

\begin{figure}[htp]
\centerline{\includegraphics[scale=0.70]{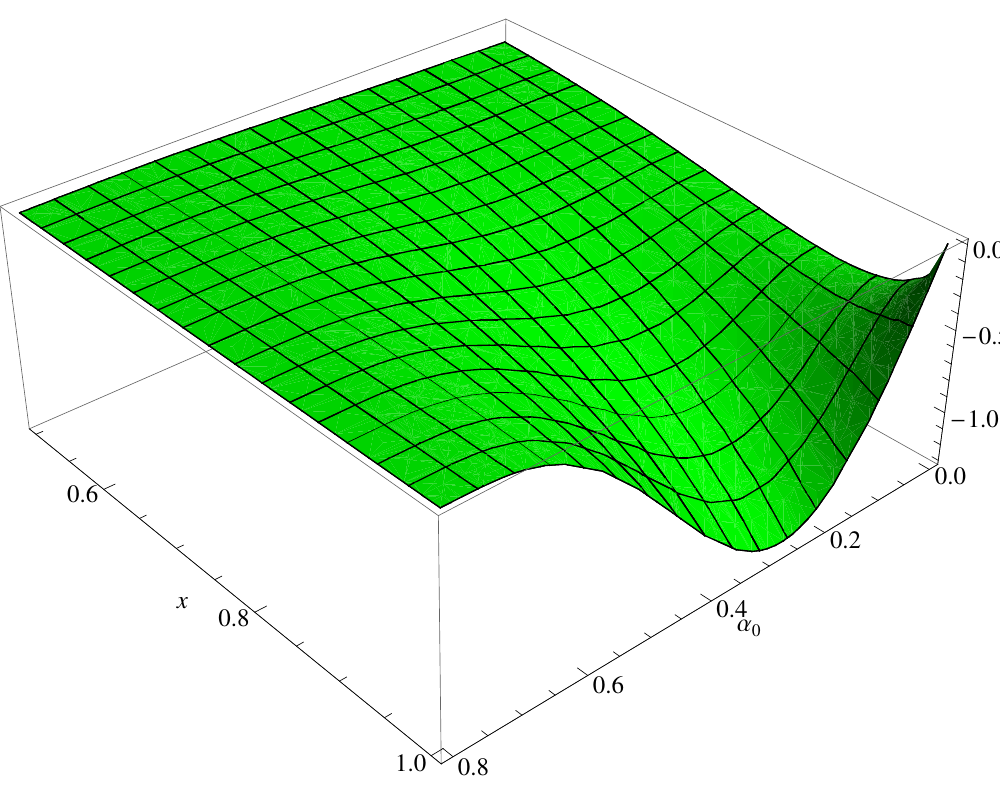}}
\vspace*{8pt}
\caption{The functional dependence $d\mathcal{C}\wedge d\mathcal{D}$ between both functions is given for all parameter combinations with vanishing values. \label{fig4}}
\end{figure}

By recalling the Schmidt-decomposed family of pure states associated to the vector 
\be \ket{\alpha_0}: = \cos(\alpha_0)\begin{pmatrix}
1 \\
0 
\end{pmatrix}\otimes \begin{pmatrix}
1 \\
0
\end{pmatrix} + \sin(\alpha_0)\begin{pmatrix}
0 \\
1 
\end{pmatrix}\otimes \begin{pmatrix}
0 \\
1
\end{pmatrix},\label{0}\ee  
having been used in the previous work for the identification of pull-back tensor fields on Schmidt-equivalence classes \cite{Aniello:09}, we may now consider a convex combination   
\be \rho_{x, \alpha_0}:= x\ket{\alpha_0} \bra{\alpha_0} +(1-x)\rho^*,\ee
with a maximal mixed state $\rho^*$. Such a convex combination contains the 1-parameter family of Werner states as special case for $\alpha_0= \pi/4$. Computing here, the purity-related invariant $f_2^R(\rho)\equiv \braket{\theta^{\rho}_R}{\theta^{\rho}_R}$ yields
\be  f_2^R(\rho_{x, \alpha_0})=f_2^R(\rho_{Werner})=  6 \left(x^2+1\right).\ee
Next we shall consider in addition, the non-linear realization, given by
\be f_2^{ \tilde{R}}(\rho)=\braket{\theta^{\rho}_{ \tilde{R}}}{\theta^{\rho}_{ \tilde{R}}}  \quad \mbox{ with }  \tilde{R}(X_j)= [R(X_j)- \rho(R(X_j))\mathds{1}]\rho\ee
to see what happens. Here we find 
$$f_2^{ \tilde{R}}(\rho_{x, \alpha_0})=2 \cos \left(4 \alpha _0\right) x^4 +\frac{1}{2} \cos \left(8 \alpha _0\right) x^4+\frac{3 x^4}{2}$$\be-2 \cos \left(4 \alpha
   _0\right) x^3-2 x^3-2 \cos \left(4 \alpha _0\right) x^2+4 x^2+6.\ee
Hence, in contrast to the Hermitian representation $R$,  the non-linear realization  $\tilde{R}$ is able to captures the information on the additional parameter $\alpha_0$. Moreover it involves a $x$-dependence beyond quadratic order. This suggests to introduce a `non-linear' extended version of the purity measure by
\be \mathcal{D}(\rho):= \frac{1}{8}f_2^{ \tilde{R}}(\rho_{x, \alpha_0})-\frac{1}{2}.\ee
This quantity shall now be compared with the concurrence $\mathcal{C}(\rho)$ by applying both functions on the considered two-parameter subclass of states $\rho_{x, \alpha_0}$. The concurrence $\mathcal{C}(\rho_{x, \alpha_0})$ yields here
\be \mbox{max}[-\frac{x^2}{8}+\frac{x}{4}+\frac{1}{2} \sqrt{-x^2 \left(2 \cos \left(4 \alpha _0\right) x^2+(x-2) x-1\right) \sin ^2\left(2
   \alpha _0\right)}-\frac{1}{8},0].\ee
The latter function is plotted in figure \ref{fig2}, while the `non-linear' extended version of the purity is given in figure \ref{fig3}.
In this regard we may compute the \emph{functional dependence} between both functions by   
\be d\mathcal{C}\wedge d\mathcal{D}=0 :\Leftrightarrow \mbox{$\mathcal{C}$ and $\mathcal{D}$ are functional dependent,}\ee
as it has been considered in \cite{Clemente:2008}. The corresponding plot is given in figure \ref{fig4}, where a region of functional independence exists.

\section{Conclusions and Outlook}\label{outlook}

In this work we have approached the separability problem of entangled mixed states $\rho$ on finite-dimensional product Hilbert spaces by means of an evaluation on invariant operator-valued tensor fields (IOVTs)
\be \theta_R :=\bigg(\prod_{a=1}^k R(X_{i_a})\bigg) \bigotimes_{a=1}^k\theta^{i_a}\ee
on the unitary group $SU(n)\times SU(n)$. By mapping the resulting invariant classical tensor fields
\be \quad \theta^{\rho}_R :=\rho\bigg(\prod_{a=1}^k R(X_{i_a})\bigg) \bigotimes_{a=1}^k\theta^{i_a}\ee
in terms of Hermitian inner product contractions to real-valued polynomial functions
$$f_k^{R}(\rho):=\sum_n a_n\braket{\theta^{\rho}_R}{\theta^{\rho}_R}^n, $$
the concept of $R$-\emph{classes} of entanglement monotones candidates emerged in dependence of the choice of a realization $R$ of the underlying Lie algebra. Let us outline in detail the two possible main practical merits of this approach for future works.\\
First, in the case of pure states, we may deal with the idea of an \emph{algorithmic} procedure to identify $SU(n)\times SU(n)$-invariant functions on Schmidt-equivalence classes (i.e. local-unitarily generated orbits) which crucially evades the computational effort of a singular value decomposition into Schmidt-coefficients. This idea might be illustrated by the fact that Schmidt-equivalence classes provide homogenous spaces \cite{Kus:2001, Kus:2002, Lupo:09}, suggesting a local, rather than a global, approach to be sufficient for extracting the information on the entanglement of the state which belongs to the corresponding orbit. Invariant covariant tensor field constructions on $SU(n)\times SU(n)$, as discussed in part I.\, in \cite{Aniello:09}, relate to pull-back tensor fields from the `quotient', i.e. from a Schmidt-equivalence class orbit and provide therefore clearly, the most natural tool for extracting this information.\\
Second, within the generalized case of mixed states, one may foresee the existence a `functorial' correspondence between certain classes of separability criteria and $R$-\emph{classes} of entanglement monotones candidates. To illustrate such a relation, we observe that we recover for the non-linear realization
\be \tilde{R}(X_j)= R(X_j)- \rho(R(X_j))\mathds{1},\ee
and the associated \emph{covariance matrix-tensor field} \be\theta^{\rho}_{\tilde{R}}=(\rho(R(X_j)R(X_k))-  \rho(R(X_j))\rho(R(X_k))\theta^j\otimes \theta^k,\ee
a class of separability criteria having their origin in the so-called \emph{covariance matrix criterion} \cite{Gittsovich:2008}. While the corresponding `linear $R$-subclass' of criteria in the Bloch-representation proposed by de Vicente \cite{deVicente:2007}, has been illustrated here in section \ref{deVicente-criteria}, we find for $n=2$, 
\be \frac{1}{8}\big(\braket{L^{\rho}_R}{L^{\rho}_R}+ (-1)^j\braket{\Omega^{\rho}_R}{\Omega^{\rho}_R}\big) -\frac{1}{2}= \begin{cases}
 \mbox{Tr}(\rho^2) &\text{ for } j=0\\
 \mbox{Tr}(\rho\tilde{\rho}) &\text{ for } j=1. 
\end{cases}\ee
and therefore the purity $\Tr(\rho^2)$, the linear entropy $1-\Tr(\rho^2)$ and the concurrence related quantity 
\be 4\Tr(\rho\widetilde{\rho})
 =1-\sum_{j=1}^3 m_j^2-\sum_{j=1}^3 n_j^2+ \sum_{j,k =1}^3 C_{j,k}^2, \label{important relation}\ee
to be in such a linear $R$-subclass according to our classification of entanglement monotone candidates.  A generalization of (\ref{important relation}) to higher dimensions may shed some light on the geometric origin of the sufficient separability criterion based on the Bloch-representation in Proposition \ref{Sufficient de Vicente}. The latter reads for two qubits
\be  \sqrt{\sum_{k=1}^3 n_j^2}+\sqrt{\sum_{k=1}^3 m_k^2} + \Tr(\sqrt{C^{\dagger}C}) \le 1\ee 
and clearly outlines a link to (\ref{important relation}). The tensorial picture may therefore allow to face the open problem (as given in \cite{Gittsovich:2008}) to relate the `non-linear $\tilde{R}$-class' of covariance matrix separability criteria to quantitative statements. In what has been said here, this could now be approached by a $\tilde{R}$-class of entanglement monotone candidates in terms of associated polynomials 
\be f_2^{\tilde{R}}(\rho)= \sum_n a_n\braket{\theta^{\rho}_{\tilde{R}}}{\theta^{\rho}_{\tilde{R}}}^n.\ee
In particular, the `first order term' $\braket{\theta^{\rho}_{\tilde{R}}}{\theta^{\rho}_{\tilde{R}}}$ indicated within the examples of a two-parameter family of entangled states in the previous section, a possible approximation to the concurrence measure.\\
The identification of non-linear structures of higher order being hidden in the description of quantum entanglement may turn out to be crucial to tackle the separability problem in both necessary and sufficient directions. This can be seen, already for the case of two qubits, by taking into account the relation between the Peres-Horodecki \emph{PPT-criterion} \cite{Peres:1996, Horodecki:1996} and the role of so-called \emph{local filtering operations} within the class of covariance matrix criteria \cite{Gittsovich:2008}.

\subsection*{The relation to the PPT-criterion}

For this purposes, let us observe, that within the linear $R$-class, we are let to consider \emph{either} sufficient \emph{or} necessary  criteria, where we found a strong indication that the anti-symmetric IOVT
$$\Omega:= [R(X_j), R(X_k)]_-\theta^j\wedge \theta^k $$
associated with the coefficients
\be \Omega^{A}_{[jk]}(\rho)=   \Tr(\rho^A[\sigma_j,\sigma_k]_-),\qquad \Omega^{B}_{[jk]}(\rho)=   \Tr(\rho^B[\sigma_{j-n^2},\sigma_{k-n^2}]_-)\ee
plays a fundamental role to measure `how far' the necessary criterion (Theorem \ref{de Vicente 1}) is from being also a sufficient criterion (Theorem \ref{de Vicente 2}). Specifically for a qubit system, we may identify the imaginary `eigenvalues' of $(\Omega^{A}_{[jk]}(\rho))$ and $(\Omega^{B}_{[jk]}(\rho))$ in dependence of the  subsystem's Bloch-representation coefficients $n_j$ and $m_k$ by
\be \pm i\sqrt{\sum_{k=1}^3 n_j^2},\qquad  \pm i\sqrt{\sum_{k=1}^3 m_k^2}.\ee
For generic entangled mixed states in $n=2$, having not maximal mixed subsystems as in the case of Werner states used in the previous section, the criterion in corollary \ref{de Vicente 1 Cor} fails to be both sufficient and necessary. A way of making this criterion stronger is given by considering the complexification of the local special unitary transformations to $SL(2,\C)\times SL(2,\C)$. In particular, as it has been shown in \cite{Leinaas:2006}, 
any state $\rho \in D(\C^2\otimes \C^2)$ can be transformed via $SL(2,\C)\times SL(2,\C)$ into the so called \emph{standard form}
\be \rho_S := \frac{1}{4}(\s_0\otimes\s_0 +d_{j} \s_j\otimes\s_j)\label{SF}\ee
with 
$(d_1,d_2, d_3)\in \triangle_2 \subset \R^3,$
parametrizing points on a 3-dimensional convex subset given by a tetrahedron, i.e. a 2-simplex $\triangle_2$. By evaluating the resulting standard form (\ref{SF}) on a symmetric IOVT
\be  L:=[R(X_j), R(X_k)]_+\theta^j\odot \theta^k\ee
associated as in the previous sections to a product representation $SU(2)\times SU(2)\rightarrow U(4)$, we find a matrix of coefficients decomposed in submatrices by 
\be (L_{(jk)})(\rho_S)=\left(\begin{array}{cc}A & C \\C & B\end{array}\right),\ee
with $A(\rho_S)=B(\rho_S)=\mathds{1}$ and
\be C(\rho_S) = \left(
\begin{array}{lll}
 d_1 & 0 & 0 \\
 0 & d_2 & 0 \\
 0 & 0 & d_3
\end{array}
\right). \ee  
By comparing with (\ref{L on rhoW2}), it becomes evident that the class of Werner states are recovered here by the constraint 
\be d_1=d_3=-d_2.\ee
In contrast to the symmetric IOVT, the evaluation on an anti-symmetric IOVT $\Omega:=[R(X_j), R(X_k)]_-\theta^j\wedge \theta^k$ will in general yield a maximal degenerate tensor field
\be (\Omega_{(jk)})(\rho_S)=0.\ee 
Now one may observe the following. A partial transposition (PT) of the matrix representation of $\rho_S$  induces a transformation on $C(\rho_S)$ given by a reflection on the second diagonal element
\be C(\rho_S) = \left(
\begin{array}{lll}
 d_1 & 0 & 0 \\
 0 & d_2 & 0 \\
 0 & 0 & d_3
\end{array}
\right) \mapsto  C(\rho_S^{PT}) = \left(
\begin{array}{lll}
 d_1 & 0 & 0 \\
 0 & -d_2 & 0 \\
 0 & 0 & d_3
\end{array}
\right). \ee 
As it has been argued in \cite{Leinaas:2006},
 the intersection
\be \triangle_2 \cap PT^*(\triangle_2) \equiv \mathcal{S} \ee
of the 2-simplex $\triangle_2$ with its PT-induced reflection $PT^*(\triangle_2)$ is identical to the set $\mathcal{S}$ of separable states, which provides an octahedron, defined by the constraint 
\be \sum_{j=1}^3|d_j|\le 1.\ee 
The later condition turns out to be equivalent to 
\be \|C(\rho_S)\|_{KF}\le 1\label{C vs PT},\ee
and crucially, generalizes the separability criterion in corollary \ref{de Vicente 1 Cor} for states with maximal mixed subsystems to generic states, whenever one makes use of a `local filtering' operation $\rho \mapsto \rho_S$ induced by $SL(2,\C)\times SL(2,\C)\subset SL(4,\C)$.\\
The latter group preserves the positive cone structure in $u^*(\C^2\otimes \C^2)\cong u^*(\C^4)$, which may allow to recover here as a consequence the Peres-Horodecki criterion \cite{Peres:1996, Horodecki:1996}, which states that a state in $D(\C^2\otimes \C^2)$ is separable iff its matrix representation remains positive under partial transposition.\\
In this regard it would be interesting to construct IOVTs, on the Lie group $\G\equiv SL(2,\C)\times SL(2,\C)$ to identify pull-back tensor fields from homogeneous space manifolds $\G/\G_0$.
The latter may become identified in this regard with distinguished strata $D^k(\H)\subset D(\H)$ of mixed states $\rho$ with fixed rank $k$, whenever one considers the non-linear action
\be \rho  \mapsto \frac{T \rho T^{\dagger}}{ \Tr(T \rho T^{\dagger})}, \quad T\in G = GL(\H)\ee
as shown in \cite{Grabowski:2000zk, Grabowski:2005my}.
By identifying the corresponding infinitesimal action of this non-linear action, in terms of vector fields resp. non-linear operators, one may build a corresponding IOVT, whose evaluation on a state with fixed rank $k$ may gives rise to a pull-back tensor from $D^k(\H)$ to $\G = GL(\H)$.\\
Such a construction could provide a tensor field which is invariant under the so called `local filtering operations' associated to the subgroup  
\be SL(2,\C) \times SL(2,\C) \subset GL(4,\C),\ee 
which transforms a given state $\rho$ to its \emph{standard form} (\ref{SF}). Invariant tensor field on these orbits could pave the way to reduce the computational effort of a local filtering operation, in analogy to invariant tensor fields on Schmidt-equivalence classes, which reduce the computational effort of a Schmidt-decomposition. This issue shall be worked out in detail, both for two qubits, but also in higher dimensional composite systems in a forthcoming paper.

\section*{Acknowledgments}
We thank D. D\"urr for discussions and encouragement on Corollary \ref{Cor L ent = max ent}.\\
This work was supported by the National Institute of Nuclear Physics (INFN).

\bibliography{mybib}
\bibliographystyle{unsrt}

\end{document}